\documentclass[useAMS,usenatbib]{mn2e}
\usepackage{graphicx}
\usepackage{natbib}
\usepackage{amsmath}
\usepackage{color}
\usepackage{arydshln}
\usepackage{tabularx}
\usepackage{colortbl}
\usepackage{hhline}

\usepackage{array}
\usepackage{multirow}

\newcommand\MyBox[2]{
  \fbox{\lower0.75cm
    \vbox to 1.0cm{\vfil
      \hbox to 2.0cm{\hfil\parbox{1.4cm}{#1\\#2}\hfil}
      \vfil}%
  }%
}

\usepackage{placeins} 

\newcommand{\url}{{url\;}}

\newcommand{\msun}{{h$^{-1}$ M$_{\odot}$\;}}

\newcommand{\mpc}{{h$^{-1}$ Mpc\;}}
\newcommand{\kpc}{{h$^{-1}$ kpc\;}}

\title[Classifying the LSS with Deep Neural Networks]{Classifying the Large Scale Structure of the Universe with Deep Neural Networks}
\author[Aragon-Calvo M.A. et al.]{M.A. Aragon-Calvo$^{1}$ \thanks{E-mail:maragon@astro.unam.mx} \\
$^{1}$Instituto de Astronom\'{i}a, UNAM, Apdo. Postal 106, Ensenada 22800, B.C., M\'{e}xico\\}
\begin{document}

\date{}

\pagerange{\pageref{firstpage}--\pageref{lastpage}} \pubyear{2002}
\maketitle
\label{firstpage}

\begin{abstract}

We present the first application of deep neural networks to the semantic segmentation of cosmological filaments and walls in the Large Scale Structure of the Universe. Our results are based on a deep Convolutional Neural Network (CNN) with a U-Net architecture trained using an existing state-of-the-art manually-guided segmentation method.
We successfully trained an tested an U-Net with a Voronoi model and an N-body simulation. The predicted segmentation masks from the Voronoi model have a Dice coefficient of 0.95 and 0.97 for filaments and mask respectively. The predicted segmentation masks from the N-body simulation have a Dice coefficient of 0.78  and 0.72 for walls and filaments respectively. The relatively lower Dice coefficient in the filament mask is the result of filaments that were predicted by the U-Net model but were not present in the original segmentation mask. 

Our results show that for a well-defined dataset such as the Voronoi model the U-Net has excellent performance. In the case of the N-body dataset the U-Net produced a filament mask of higher quality than the segmentation mask obtained from a state-of-the art method. The U-Net performs better than the method used to train it, being able to find even the tenuous filaments that the manually-guided segmentation failed to identify.

The U-Net presented here can process a $512^3$ volume in a few minutes and without the need of complex pre-processing. Deep CNN have great potential as an efficient and accurate analysis tool for the next generation large-volume computer N-body simulations and galaxy surveys.

\end{abstract}

\section{Introduction}\label{sec:intro}

The advances in artificial intelligence in the last decade have resulted in artificial neural networks that outperform existing methods in virtually any area of application \citep{LeCun15}. Specifically deep neural networks (DNN) and convolutional neural networks (CNN) are propelling the current revolution in image processing / pattern recognition. The ability of deep neural networks to create a hierarchical internal representation of features allows them to identify complex patterns where traditional hand-crafted methods fail. 

One of the main challenges in the study of the Large Scale Structure (LSS) of the Universe is the identification and segmentation of its basic geometric components: spherical clusters, elongated filaments and planar walls, with voids being the empty spaces delineated by these structures. One of the reasons for this difficulty is the intrinsic complexity of the LSS: the basic morphological components of the LSS  form an interconnected network covering a wide range of densities and have a multiscale and hierarchical nature: voids are not empty but contain a tenuous network resembling a small-scale version of the LSS \citep{Aragon10b, Aragon13a} and walls are criss-crossed by a net of smaller filaments \citep{Hahn07, Aragon14b} which are themselves punctuated by haloes. 

\begin{figure*}
  \centering
  \includegraphics[width=0.8\textwidth,angle=0.0]{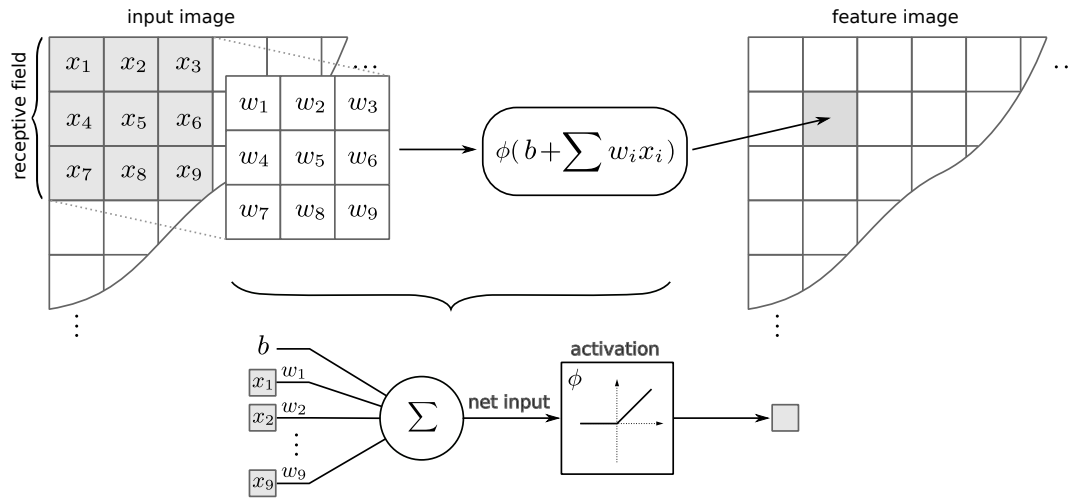}
  \caption{ Top: convolutional Neuron applied to a 2D image. The input signal corresponding to pixels $x_1, x_2 ... x_9$ are multiplied by the weights $w_1,w_2,... w_9$ in the filter and a bias term is added to produce the net input $z$. The net input is then evaluated with an activation function $\phi$ to produce a single output pixel in the feature image. The same operation is applied over all the pixels in the input image in a convolution fashion. Bottom: stylized representation of the top sub-figure showing the input, weights, bias, net input and activation function in the more familiar artificial neuron illustration. Note that the neuron is centered in pixel $x_5$.  For simplicity we use indexes in the reference frame of the filter instead of the input image.}\label{fig:cnn}
\end{figure*}

%
\subsection{Feature engineering}

The field of LSS classification\footnote{The labeling of regions of space belonging to a certain class of objects (semantic segmentation) can be considered a pixel-based classification problem that includes the pixel's context.}  has been dominated by techniques based on \textit{feature engineering}, i.e. the design of specific filters or criteria tuned to identify particular patterns in the data. These methods include the percolation properties of the galaxy distribution and point processes \citep{Barrow85,Stoica05,Tempel16}, the geometry of the density field identified by a set of morphology filters or geometric criteria  \citep{Babul92,Aragon07b,Hahn07},  local dynamics \citep{Forero-Romero09,Hoffman12,Cautun13,Fisher16} and the topology of the density field \citep{Sousbie08,Aragon10a,Neyrinck12} to name a few (see \citet{Libeskind18} for a recent comparison between methods). By designing specific filters one can have full control over the characteristics of the structures we seek to identify as well as a good understanding of the nature of the filter itself. However, this comes at the cost of poor generalization and limited applicability as the filters are often designed following a narrow criteria. In addition to that some methods are computationally expensive and can become prohibitive for observations/simulations spanning a large volume.

%
\subsection{Machine learning}

A different approach to feature engineering for pattern recognition problems is \textit{machine learning} in which the data itself is used to train a model and create an internal representation of the features we want to identify. A machine learning system consist of a general model, usually with a large number of parameters, that can be trained, or tuned, to perform a given task. The training is done by evaluating the model using a data sample for which we know the desired result. The difference between the response from the model and the desired response is used to update the parameters of the model in an iterative way using gradient-based optimization algorithms  \citep{Kingma14} \footnote{For clarity the above description is over simplistic and  ignores practical challenges. For an introduction to machine learning see \citet{Duda01a}.}. 

\begin{figure*}
  \centering
  \includegraphics[width=0.99\textwidth,angle=0.0]{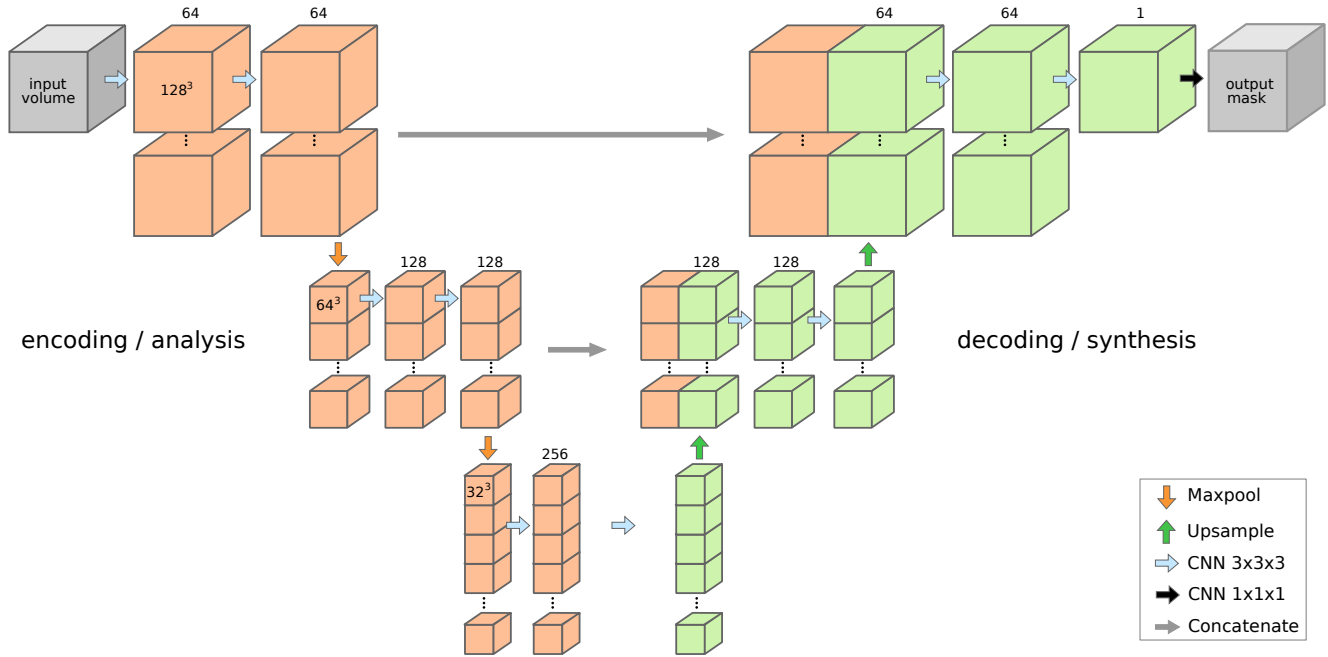}
  \caption{Basic U-Net architecture with resolution 3-levels. Cubes represent tridimensional feature maps colored orange for the encoding/analysis path and green for the decoding/synthesis path. Colored arrows indicate  operations performed by different layers and connections between layers (see legend in the figure).  Horizontal gray arrows indicate merging of feature maps used to transfer small-scale spatial information from the encoding/analysis layers into the decoding/synthesis layers. The number of filters in each layer is indicated at the bottom of each column and the size of the cubes is shown inside the top cube of each column. Note that the number of feature images increases with the number of filters. The U-Net without the the merging links results in an autoencoder network.}\label{fig:UNet}
\end{figure*}

%
\subsection{Convolutional neural networks}

One of the most effective machine learning tools are artificial neural networks which, based on a loose analogy with biological systems, form interconnected networks of simple neurons that combine to learn complex patterns. An artificial neuron takes a set of inputs $x_i$ and multiplies them by a set of trainable weights $w_i$ plus a bias factor $b$ to compute the so-called net input:

\begin{equation}
  z = \sum w_i x_i + b
\end{equation}

\noindent The net input is evaluated by an activation function which quantifies its significance. In practice it has been found that the non-linear function Rectified Linear Unit (ReLU) produces good results and has the advantage of being computationally less expensive than other non-linear activation functions such as the sigmoid and hyperbolic tangent functions \citep{Glorot10,Andrew13}. The ReLU activation function is defined as:

\begin{equation*}
    \phi(z) = 
    \begin{cases}
      z & \text{if } z \ge 0,\\
      0 & \text{if } z < 0.
    \end{cases}
\end{equation*}

For problems where spatial information is important a particular type of neural networks called \textit{Convolutional Neural Networks} (CNN) can be used to learn spatial features (see Fig. \ref{fig:cnn}). In a CNN a neuron (also called kernel or filter) is applied to each pixel and its neighborhood as it sweeps through the input image\footnote{In what follows we use the term image to denote a 2D or 3D array and pixel to denote pixel or voxel depending on context.} (hence the term convolutional) to produce a  \textit{feature image} containing high values in pixels matching the pattern encoded in the weights and bias of the neuron. A convolutional layer typically contains several filters in order to encode different aspects of the input image.

%
\subsection{Deep Neural Networks}

A single layer of neurons has limited learning ability so in practice several layers are stacked forming a \textit{Deep Neural Network} (DNN). Such neural networks have the ability to learn increasingly complex patterns as more layers are added. As the information moves across different layers of neurons the raw information in the input image is transformed into an internal representation that encodes the most relevant features of the image. In a deep neural network the feature images produced by a convolutional layer become the input images of the next layer.

%
\subsection{The U-Net architecture}

One of the basic neural network architectures for data encoding/decoding (a fundamental task in segmentation) is the autoencoder  (see \citet{Hinton504}) which consists of an encoding path with decreasing spatial resolution followed by a decoding path with increasing spatial resolution. This \textit{bottleneck} configuration forces the autoencoder to learn an internal compressed representation of the input image as the information travels down the encoding path. The image is then reconstructed as the spatial resolution increases going up the decoding path. Autoencoders have, in general, limited practical utility as they suffer from poor locality. In order to increase locality in the autoencoder architecture for semantic segmentations problems \citet{Ronneberger15} proposed the U-Net architecture in which the small-scale features from the encoding path are used to add detail in the decoding path. The U-Net architecture can produce detailed and accurate semantic segmentations and has been successfully applied over a wide range of imaging problems becoming one of the standard architectures for semantic segmentation. The U-net, originally proposed for 2D medical imaging segmentation, has recently been applied to the segmentation of 3D datacubes to identify structures of medical interest \citep{Cicek16, Milletari16, Casamitjana17}.

Figure \ref{fig:UNet} shows the U-Net architecture with 3 main levels (we define a level as a row in the U-Net architecture with data cubes of the same spatial resolution). The cubes represent feature layers resulting from the application of operators (CNN, maxpool, concatenation, upscaling). As the spatial resolution decreases along the encoding path, the number of filters increases in order to store the internal representation of increasingly complex patterns. Going up the decoding path the spatial resolution increases and the number of filters decreases. In order to add the small-scale features needed for spatially accurate semantic segmentation the layers in the decoding path are connected to the encoding path. The need for transfer of information between layers at opposite sides of the encoding/decoding path can be seen in Fig. \ref{fig:UNet} where at the bottom of the network the feature images have a size of $32^3$ voxels. At this point the features in the image are encoded in the convolutional filters but they have poor spatial resolution. It is impossible to recover the original location of the small-scale features in the input $512^3$ image from the $32^3$ feature images. By using information from the encoding layers we can add the required details into the decoding path.

%
\section{Implementing the U-Net}\label{sec:unet}

Figure \ref{fig:UNet} shows the U-Net architecture used in this work. The particular U-Net network we implemented consists of a series of  \textit{bottleneck stacks} containing two convolutional layers with Rectified Linear Unit (ReLU) activations followed by a maxpooling layer decreasing the spatial resolution by half.  Each convolutional layer has filter kernels of size $3\times 3\times 3$ and stride of $1 \times 1 \times 1$ except for the upsampling layer (also refereed to as \textit{deconvolution}) which has a kernel of size $2\times 2\times 2$ and stride of size $2\times 2\times 2$\footnote{the fixed properties of the network are refereed to as hyperparameters}.  The borders were treated using a padding that resulted in an output with the same size as the input image.

The kernel size, stride and the number of stacked layers determines the extent of the \textit{receptive field} which limits the size of the largest features that will be processed and learned by the network. As such it must be large enough to fully include the structures we want to identify and also the necessary context for segmentation. A small receptive field will focus on details such as filament/wall geometry and density profiles while a large receptive field will include cosmological context such as the interconnectivity of the LSS. A network with a receptive field smaller than the typical thickness of a filament/wall will not be able to learn their geometry and will learn less optimal features. A large receptive field that encloses complete voids will learn (if it is also deep enough) not only the geometry of filaments/walls but perhaps their interconnectivity. However this will require a much larger number of filters (and training samples) to learn all the possible variations in the LSS. 
In our case we are interested in producing a segmentation of filaments and walls on a voxel-by-voxel basis which requires limited contextual information. 
The 4-level U-Net implemented in this work has kernels of size $3 \times 3 \times 3$ and stride of $1 \times\ 1 \times 1$. This results in a receptive field of 52 voxels or 6.5 \kpc at the bottom of the network, which is enough to cover typical filaments and walls.


The number of filters in each convolutional layer determines the amount of information the network will be able encode. The first layer learns simple gradient-based features so in principle a small number of filters is necessary (however the filters must be able to learn the tridimensional orientations of the features). We found that 64 filters in the first layer are enough for our purposes. This is a lower limit since we have no control on the shape and orientation of the learned filters and CNNs can generate redundant filters. The U-Net network we implemented has 4 levels with 64,\;128,\;256 and 512  filters of size $(3\times3\times3)$ resulting in 22,393,793 trainable parameters (see table \ref{tab:layers}).


We quantify the accuracy of the predicted semantic segmentation mask using the Dice coefficient \citep{Dice45} defined as:

\begin{equation}
\textrm{Dice}(X,Y) = \frac{ 2 | X  \cap Y|}{ |X| + |Y| }
\end{equation}

\noindent where $X$ and $Y$ are the set of voxels inside the segmentation mask in the training and predicted data respectively. The Dice coefficient indicates the degree of similarity between two sets (in our case the voxels predicted for the class). It is similar to the intersection-over-union measure (also know as Jaccard index) but uses the sum of the two sets instead of their union. 

\begin{table}

  \begin{center}
    \leavevmode

\begin{tabular}{l l c c c} 
\hline
\hline
Name                 & type        & filters & image size & parameters\\
\hline 
encode-1a  & Convolution  & 64     & $128^3$  & 1,792\\
encode-1b &Convolution    & 64   & $128^3$   & 110,656\\
pool-1           & Maxpool & & $64^3$   & \\[0.05in]
encode-2a & Convolution  & 128     & $64^3$ & 221,312\\
encode-2b & Convolution  & 128     & $64^3$  & 442,496\\
pool-2           & Maxpool & & $32^3$  & \\[0.05in]

encode-3a & Convolution   & 256    & $32^3$ & 884,992 \\
encode-3b & Convolution    & 256    & $32^3$ &  1,769,728 \\
pool-3           & Maxpool &  & $16^3$  & \\[0.05in]

bottom-a & Convolution     & 512  & $16^3$  & 3,539,456\\
bottom-b & Convolution    & 512  & $16^3$ & 7,078,400 \\[0.05in]

up-3 & Deconvolution         &  & $32^3$   & 1,048,832\\
decode-3a & Convolution     & 256  & $32^3$  & 3,539,200\\
decode-3b & Convolution     & 256 & $32^3$  & 1,769,728 \\[0.05in]

up-2 & Deconvolution         &  & $64^3$   & 262,272\\
decode-2a & Convolution     & 128  & $64^3$  & 884,864\\
decode-2b & Convolution     & 128 & $64^3$  & 442,496 \\[0.05in]

up-1 & Deconvolution         &  & $128^3$  & 65,600 \\
decode-1a & Convolution     & 64  & $128^3$  & 221,248 \\
decode-1b & Convolution      & 64  & $128^3$ & 110,656 \\[0.05in]

output & Convolution       & 1 & $128^3$  & 65\\

\hline 
\hline
  \end{tabular} 

\caption{Layers in the 4-level U-Net architecture. All CNN layers have kernels of size $3 \times 3 \times 3$. The deconvolution layers combine upsampling and concatenation between layers in the decoding and encoding paths. Note that while we include the image size at each layer the network architecture does not depend on the input size.}
  \label{tab:layers}
  \end{center}

\end{table}

%
%
%
%

\subsection{Practical implementation}

We implemented the U-Net described in this section using the the Keras\footnote{https://keras.io} library\citep{chollet2015keras} with the TensorFlow backend \citep{tensorflow2015-whitepaper}. Table \ref{tab:layers} shows the individual layers in our U-Net implementation. Since the U-Net we implemented contains only convolutional, maxpooling and upsampling layers we do not need to specify an input size and so the model only contains information on the trainable parameters making it possible to predict any input volume as long as it can be consistently downsampled by the 3 bottleneck stacks. Training and prediction was done on a dedicated server with an Nvidia Titan-X GPU card.

%
\section{Generating training data}

The generation of training data is a crucial step in the training of a machine learning model as from it depends the quality of the resulting predictions. In artificial vision and medical imaging the structures of interest for segmentation often have well defined and sharp boundaries and can be objectively labeled (usually this task is done by hand). In the case of cosmic structures the picture is more complex. While haloes can be defined using a physically-motivated density threshold (making them easy to identify), in the case of filaments and walls there is not single threshold that can be used to differentiate between them (see for instance  Fig. 11 in \citet{Aragon10a}). 

Ideally the training data for semantic segmentation must  have i) high accuracy, ii) unique solution, iii) self consistency and iv) high degree of variation. For the problem at hand this means that the segmentation used for training must include all the filaments/walls in the density field while having a low rate of false positives/negatives. Every voxel in the training segmentation must have one single label. The definition of filaments/walls must be consistent across the dataset. If overlapping of conflicting criteria are used for segmentation this will affect the training process as the network may end up chasing multiple minima for the same class. Finally, the data must have a high degree of variation in order to increase the generalization ability of the network and to prevent the network from memorizing the actual training features (this is called overfitting and it is an important problem to consider when training neural networks). 

In the rest of this section we describe the procedure we followed to generate training data using a Voronoi model and an N-body simulation.

\begin{figure*}
  \centering
  \includegraphics[width=0.9\textwidth,angle=0.0]{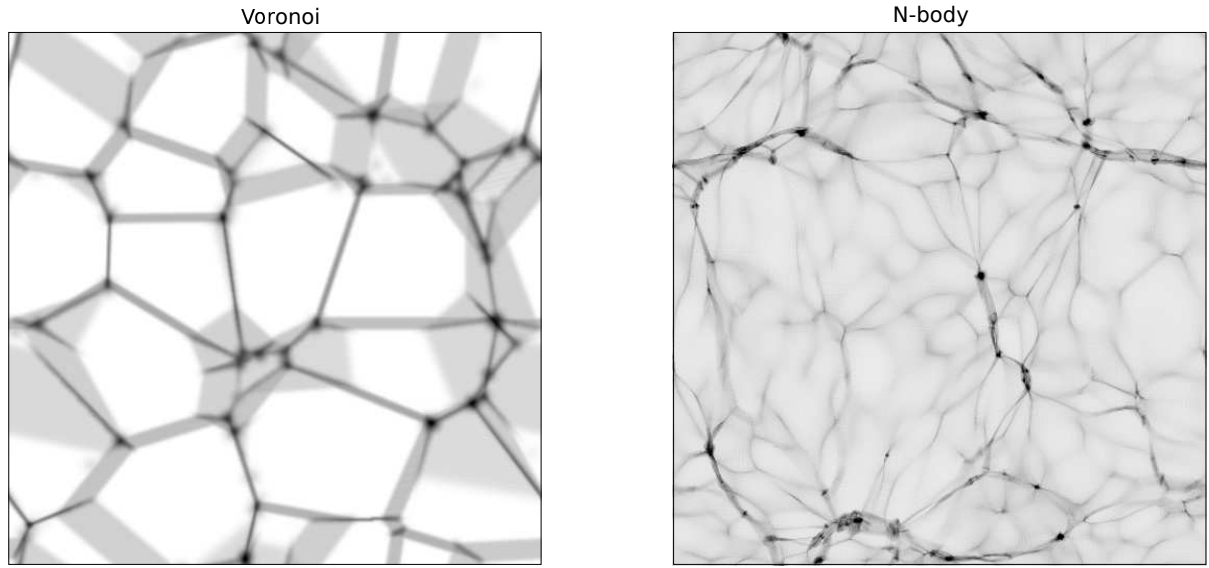}
  \caption{Input images used in this work. Left: Voronoi model showing the three main components: nodes, filaments and walls constructed as described in Sec. \ref{sec:Voronoi}. Right: Density field computed from the  N-body simulation (Sec. \ref{sec:N-body}). The intensity of the density field has been scaled to reduce its dynamic range.}\label{fig:models}
\end{figure*}

\subsection{Dataset I: Voronoi models}\label{sec:Voronoi}

In order to quantitatively measure the performance of the proposed U-Net we created a heuristic model that resembles the LSS and its  main morphological components. We implemented a Voronoi model \citep{Icke91,Okabe00} in which nodes, faces and edges of the tessellation correspond to clusters, filaments and walls respectively as done in \citet{Aragon10b} (see also \citet{Neyrinck17}). The Voronoi model allows us to have full control of the simulated elements of the LSS and generate an objective semantic segmentation.

We created a box of 100\mpc of side containing several Voronoi seeds with a number density similar to the void number density found in simulations \citep{Platen08}. Vertices, edges and faces of the Voronoi tessellation were independently sampled on a regular grid of $512^3$ voxels producing three fields with one-pixel thick clusters, filaments and walls respectively. Each field was smoothed with a Gaussian kernel of 1.5 1  and 0.5 \mpc, roughly corresponding to the scale of typical cosmic structures observed in the galaxy distribution.  The intensity of each field was then scaled to a maximum value of 10,5 and 1, following the increasing density of clusters, filaments and walls respectively. Finally the three fields were added together (see Fig. \ref{fig:models}). 

The segmentation masks used for training and to compare with the predictions from the network were created by applying a threshold corresponding to three sigmas (this was done independently for each morphological component). 

\subsection{Dataset II:N-body simulation}\label{sec:N-body}

The cosmic web traced by luminous galaxies of particles in a computer simulation is far more complex than the Voronoi models described in the previous section. The multiscale and hierarchical nature of the LSS as well as the interconnectivity of its elements makes it unfeasible to fully describe a given region of space in terms of a single geometry (see section \ref{sec:intro}). One is the faced with the challenge of producing an objective and accurate  semantic segmentation from an ill-defined problem. One way to address this is to restrict the problem to a better defined space. In the case of N-body simulations we can break the hierarchy of the LSS and avoid ill-defined cases with ambiguous morphology such as intra-wall filaments and and intra-filament nodes by applying a low-pass filter at the initial conditions (before Fourier mode coupling occurs) to target cosmic structures in a well defined scale range  \citep{Aragon10b,Aragon13a,Aragon14b}.

The N-body training and testing data used in this work was obtained from an N-body cosmological simulation run inside a 64 \mpc box, assuming a $\Lambda$CDM cosmology with values of $\Omega_m=0.3$, $\Omega_{\Lambda}=0.7$, $h=0.73$ and $\sigma_8=0.8$, similar to the latest values obtained from the Planck mission \citep{Planck15}. The initial conditions were smoothed with a Gaussian kernel of 2 \mpc as described in \citep{Aragon14b}. The simulation was run until $z=0$ using the Gadget-2 code \citep{Springel03}.

%
\subsubsection{Density field estimation}\label{sec:LTFE}

We computed a continuous density field from the discrete particle distribution using the Lagrangian sheet method \citep{Shandarin12,Abel12}. In this approach matter is assigned to tetrahedral volume elements defined by the initial particle grid. As particles evolve density changes with the deformation of the Lagrangian volume element (assuming mass conservation inside each tetrahedra). The density field inside a regular grid is obtained by sampling the tetrahedral tessellation into a grid. In order to minimize artifacts introduced by the choice of the tetrahedra orientation we estimate densities for the 8 possible tetrahedral decompositions of a cube and compute their mean (while this does not conserve mass it produces density fields with less tetrahedral artifacts). The Lagrangian sheet method produces the best available density estimation in the low to intermediate density ranges ($0 < \delta < 10 $) and can be efficiently computed.

%
\subsubsection{Filament/Wall segmentation}\label{sec:MMF}

The semantic segmentation of the density field was computed using the latest implementation of the Multiscale Morphology Filter (MMF, \citet{Aragon07b,Aragon14b}) based on the use of Lagrangian-smoothed initial conditions to target specific scales in the cosmic web hierarchy. The identification of structures in the MMF method is based on the second-order local variations of the density field encoded in the Hessian matrix ($\partial^2  \rho/ \partial x_i \partial x_j$). LSS geometries are described in terms of ratios between the eigenvalues of the Hessian matrix ($\lambda_1 < \lambda_2 < \lambda_3$). Different  combinations of ratios measure local spherical symmetry, filamentariness, or planarity. The ratios are then thresholded to produce a segmentation mask of voxels identified as either belonging to a given LSS morphology. 

The determination of the optimal threshold is a crucial step of the MMF method, a low value will produce a mask including most filaments but it will also introduce noisy structures while a high value will result in cleaner masks but will only include a few prominent structures. In  \citet{Aragon07b} we proposed an automated algorithm based on the connectivity of the cosmic web. While the method produces good segmentations often it is possible to produce a better mask by fine-tuning the threshold after visual inspection. In order to produce the best possible segmentation we selected the optimal threshold in a semi-automated way: first we used the automated algorithm described in \citet{Aragon07b} and then refined it by visually inspecting a range of threshold values around the one found in the first step. Although this introduces a subjective criteria (informed by years of experience in LSS classification by the author) the final threshold is simply a fine-tuned value close to the one proposed by the automated algorithm. Finally, the filament/wall masks were visually inspected again and small/noisy structures remaining in the masks were removed. 

We computed the wall, filament and cluster (see below) masks in a sequential way. In order to enforce uniqueness in the segmentation masks we removed voxels labeled as clusters from the filament mask and voxels labeled as filaments+nodes from the wall mask as follows:

\begin{equation}\label{eq:fila_clean}
F_{\textrm{\tiny{clean}}} = F - C
\end{equation}

\begin{equation}\label{eq:wall_clean}
W_{\textrm{\tiny{clean}}} = W - (F \cup C)
\end{equation}

\noindent where $F$ is the filament mask, $C$ is the cluster mask and $W$ is the wall mask. The cluster mask was computed by sampling all particles inside haloes more massive than $5\times10^{12}$ \msun. This mass is a lower limit for large galaxies and small groups that can be found at the intersection of filaments. Haloes were identified using the friends-of-friends group finder with a linking length of $b=0.2$ times the mean interparticle separation.

In the present work we do not focus on the segmentation of clusters since they are straightforward to identify with particle-based methods. Also, being compact objects their segmentation would be strongly dependent on the grid resolution using the approach presented here. Filaments and walls, on the other hand, are extended objects that can be better sampled by the grid.

\begin{figure*}
  \centering
  \includegraphics[width=0.9\textwidth,angle=0.0]{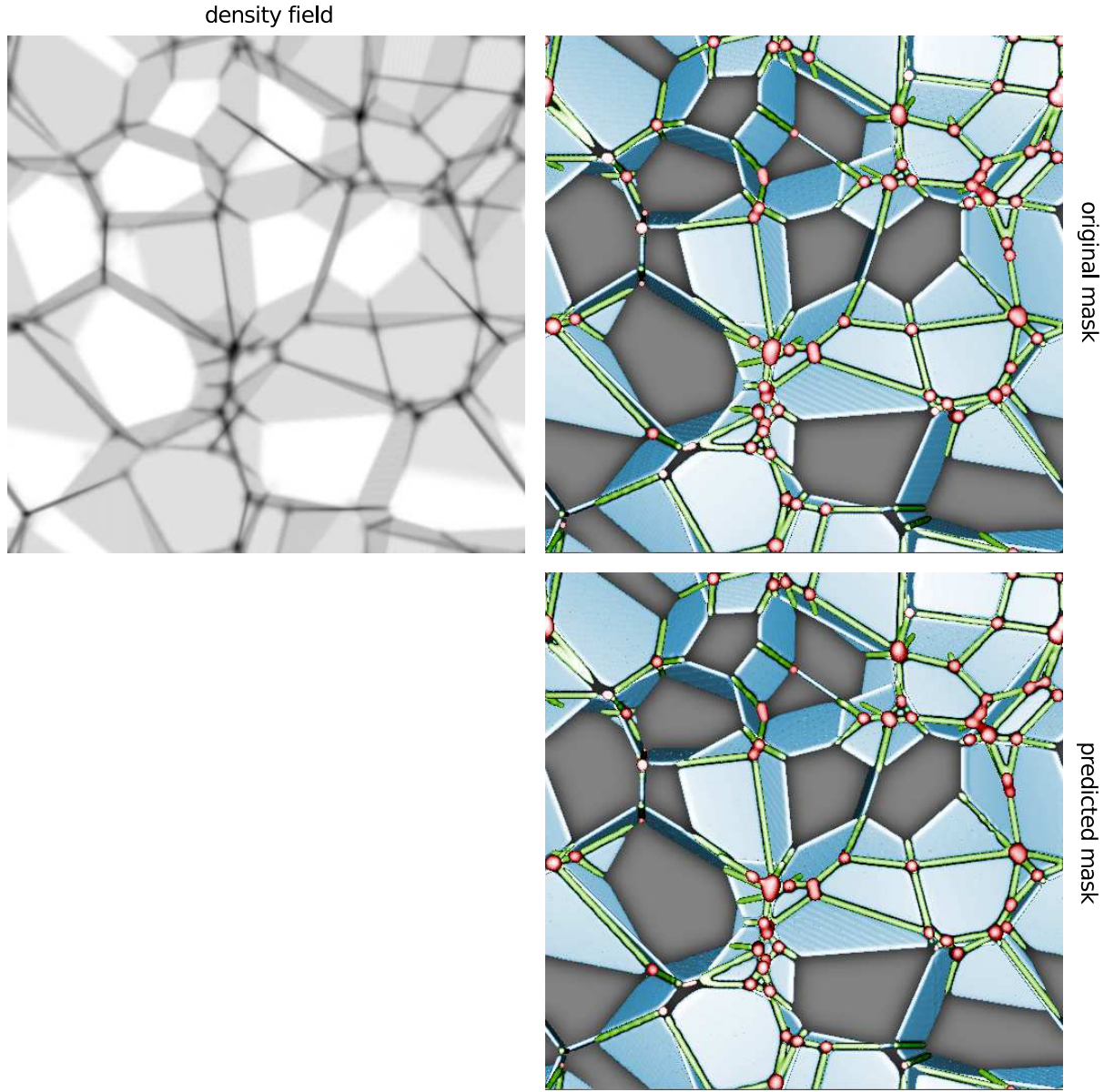}
  \caption{Filament and wall segmentation of a Voronoi model. Top left: input density field constructed by sampling the Voronoi model on a regular grid (see sec. \ref{sec:Voronoi} for details). Top-right: original wall (blue), filament (green) and cluster (red) segmentation masks. Bottom-right: predicted segmentations masks with each component colored as in the original mask. The semantic segmentation was performed on each component independently and combined in one single figure. The individual (binary) masks were slightly smoothed for illustration purposes.}\label{fig:voronoi_prediction}
\end{figure*} 

%
\section{Results}

In this section we present the results of training and applying the U-Net described in section \ref{sec:unet} to the semantic segmentation of filaments and walls from the Voronoi model and N-body simulation (sections \ref{sec:Voronoi} and \ref{sec:N-body}). Our main results are presented in Figures \ref{fig:voronoi_prediction}, \ref{fig:64Mpc_prediction_fila} and \ref{fig:64Mpc_prediction_wall}.

%
\subsection{Training}

We trained the U-Net on both the Voronoi and N-Body models using a grid size of  $512^3$ voxels. The input cube was divided into smaller non-overlapping sub-cubes of $128^3$ voxels each. In order to reduce overfitting by providing the network with a larger training sample we implemented a simple data augmentation technique and rotated the sub-cubes over each axis  by $90^{\circ}$ producing a total of 4 versions (the original and three rotations). This resulted in 256  training sub-cubes. The array of training sub-cubes was shuffled prior to training in order to avoid feeding the U-Net consecutive rotated cubes. The intensity of the voxels inside each sub-cube was normalized in the range $[0-1]$. This removes some of the information encoded in the intrinsic filament/wall density distribution, which in principle could be used by the U-Net, but also has the effect of forcing the model to put more emphasis on local geometry instead of local density.
Each U-Net was trained for 30 epochs with 204 samples for training and 52 samples for testing/validation (total sample was divided into training and testing at $80\%$ and $20\%$ respectively) and a batch size of 1 sample per epoch (determined by GPU memory constraints). Training took $\sim400$ seconds per epoch ($\sim$ 2 sec. / step). The models were trained using Adam optimization with a fixed learning rate of $10^{-5}$. For loss metric we used the negative of the Dice coefficient.

%
\subsection{Prediction}

We tested the U-Net with a $45$ degree rotated version of the training datacubes. Since the network was trained from orthogonal versions of the original volume this rotated version is sufficiently different to be used for testing.  Due to GPU memory limitations it is not possible to feed the U-Net with the complete $512^3$ cubes for training. Instead we divided the cube into sub-cubes of $128^3$ voxels in order to minimize border effects by maximizing the volume/area ratio while staying within memory constraints. 

Since the receptive field of the network at the boundary of the sub-cubes can only see partial structures the prediction in those regions tends to give poor results. To correct for this we computed the segmentation on sliding sub-cubes of $128$ voxels of side and $64$ voxels of stride on each dimension and removed the overlapping 32 voxels at the boundaries.

%
\subsection{Voronoi models}

Figure \ref{fig:voronoi_prediction} shows the predicted mask for clusters, filaments and walls in the Voronoi model. The training reached an Dice coefficient of $\sim  0.9$ for both filaments and walls after just three epochs and 0.95 and 0.97 for filaments and walls respectively after 30 epochs. The simple geometry of the structures makes the U-Net reach a high accuracy very early in the training process. The only difference between individual structures of the same morphology in the Voronoi model is their orientation, making this a very simple problem for the U-Net. Despite the apparent lack of variation in the training data the U-Net does not seems to overfit (or overfits very lightly) as both training and testing losses remain very similar throughout the 30 training epochs.  

Both the original and predicted masks in Fig. \ref{fig:voronoi_prediction} are practically indistinguishable as indicated by the high Dice coefficient. 
Our results show that for a well defined dataset the U-Net produces a nearly perfect segmentation. However, given the simple geometry of clusters, filaments and walls in the Voronoi model these results must be taken only as best-case scenario. For such dataset a simpler U-Net may be sufficient to capture all the relevant features needed for segmentation. Still, the results obtained here are on par with classic state-of-the-art algorithms for LSS segmentation (see for instance Fig. 5 in \citet{Aragon10a}). 

\begin{figure}
  \centering
  \includegraphics[width=0.45\textwidth,angle=0.0]{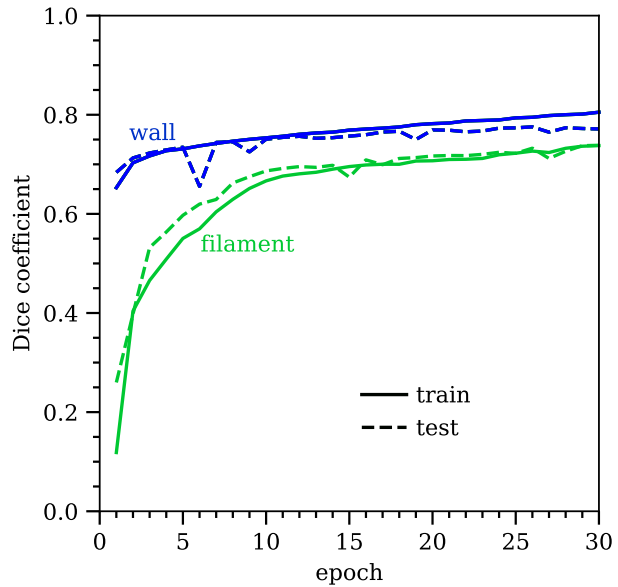}
  \caption{Dice coefficient as a function of training epoch for the filament (green) and wall (blue) models. Solid lines show the Dice coefficient  computed during  training and dashed lines during testing. The point where the training and testing lines cross (around epoch 25 and 10 for filaments and walls respectively) indicates the epoch at which the network starts to overfit.}\label{fig:64Mpc_train_test}
\end{figure} 

\begin{figure}
  \centering
  \includegraphics[width=0.45\textwidth,angle=0.0]{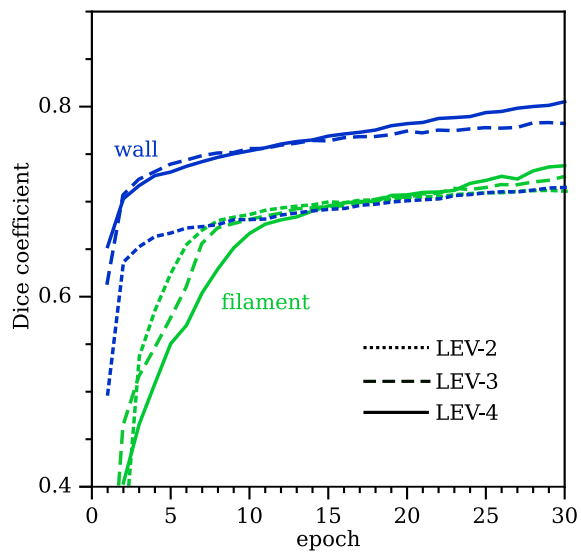}
  \caption{Dice coefficient as a function of training epoch computed during training of the U-Net architecture with 2 levels (LEV-2, dotted line), 3 levels (LEV-3, dashed line) and 4 levels (LEV-4 solid line). Green lines correspond to the filament model and blue lines to the wall model. For clarity we only included the Dice coefficient computed during training. Note the different range in the y axis compared to Fig.  \ref{fig:64Mpc_train_test}}\label{fig:64Mpc_LEV3_vs_LEV4}
\end{figure}

%
\subsection{N-body simulation}

This section presents the results of training the U-Net network on the final snapshot ($z=0$) of the N-body simulation described in sec. \ref{sec:N-body}. Since the density field at $z=0$ has a log-normal distribution \citep{Coles91,Neyrinck09} we applied two simple pre-processing steps to reduce its dynamic range and facilitate training. First we saturated all values with $\delta+1 > 100$ and then applied a quadratic root function (we used this approach instead of a logarithmic function since the logarithm tends to enhance very tenuous noisy  structures inside voids). The filament and wall segmentation masks used for training were computed with the manually-guided MMF-2 method  described in section \ref{sec:MMF}. 

\begin{figure*}
  \centering
  \includegraphics[width=0.99\textwidth,angle=0.0]{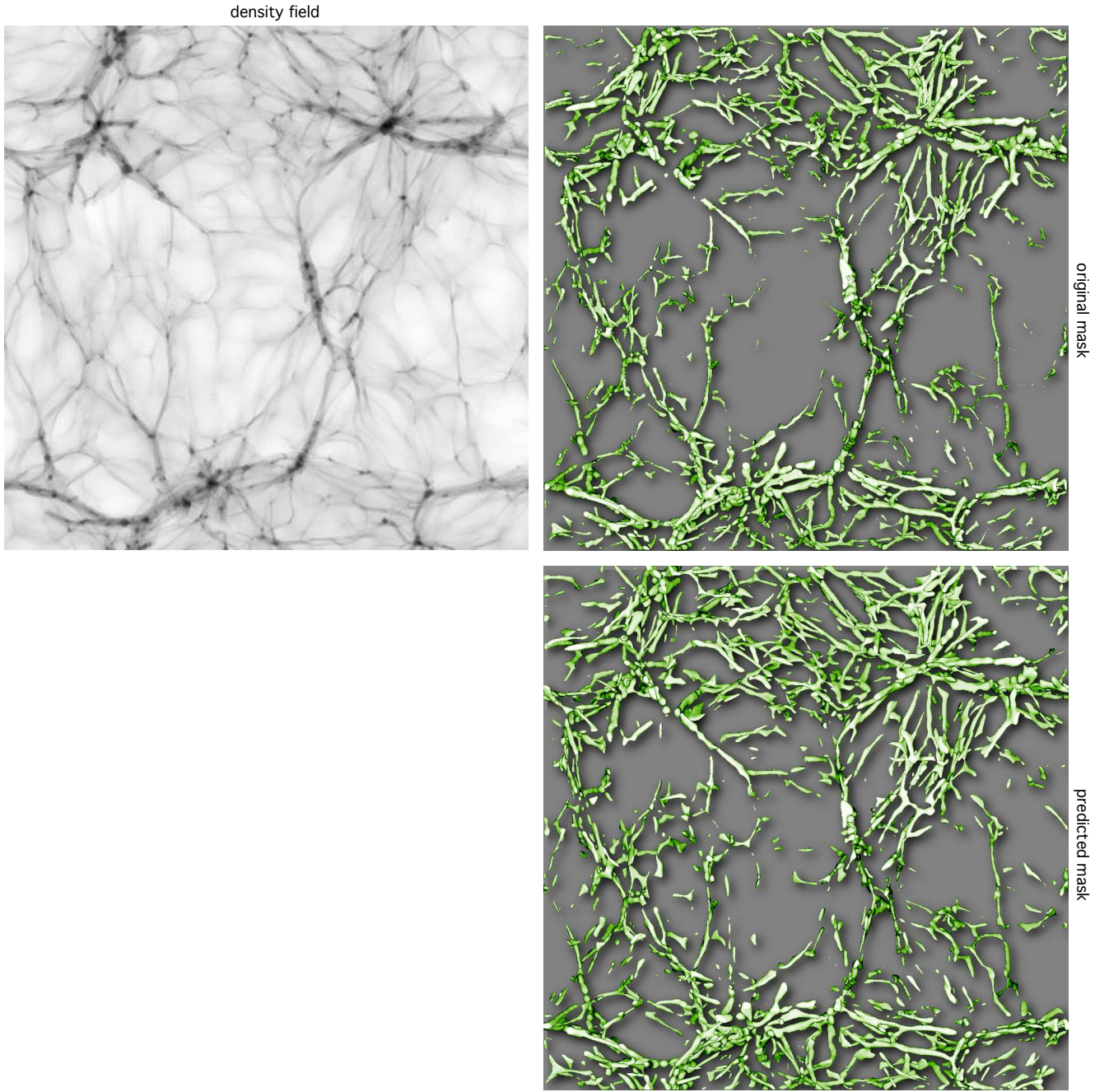}
  \caption{Filament segmentation of the N-body model. Top left: input density field. Top-right: original segmentation mask computed with the MMF method after manually-guided thresholding and cleaning. Bottom-right: predicted segmentation mask computed with the LEV-4 U-Net model. Gray background and shading added for depth cues.  The (binary) masks were slightly smoothed for illustration purposes.  Note that there are more filamentary structures in the predicted mask compared to the MMF mask.}\label{fig:64Mpc_prediction_fila}
\end{figure*}

Figure \ref{fig:64Mpc_train_test} shows the Dice coefficient computed during training for the filament and wall models as a function of the number of training epochs. The Dice coefficient increases rapidly for the first 3 and 10 epochs for walls and filaments respectively, and then reaches a regime of slow increase. After 30 epochs the Dice coefficient is $0.72$ and $0.78$ for filaments and walls respectively. The filament model overfits around epoch 25 and the wall model overfits around epoch 10. Note that the values of the Dice coefficient reported during training are slightly lower (a few percent) than the actual values since the segmentation fails at the boundaries of the sub-cubes. 

%
\subsection{Effect of network depth}\label{sec:effect_depth}

Figure \ref{fig:64Mpc_LEV3_vs_LEV4} shows a comparison between the Dice coefficient computed during training of the U-Net with 4 levels used in this work (LEV-4) and two shallower U-Net architectures with 2 levels (LEV-2) and 3 levels (LEV-3). The U-Net with 4 levels has a slightly better performance compared with LEV2 and LEV3 but at a higher memory cost. The LEV-4 models has 22,393,793 parameters while the LEV-3 model has 11,873,985 parameters and LEV-1 has 8,670,273 parameters. This is in addition to the larger number of feature images needed for additional levels. Computing time was similar for all models. The LEV-2 network has a similar performance than LEV-3 and LEV-4 in the case of filament segmentation but a poor performance for wall segmentation. It is interesting to note that, for filament segmentation, shallower networks learn faster than deeper networks but reach a lower performance at the end of the training. 

Given the small difference in performance between the LEV-3 and LEV-4 networks for both filament and wall segmentation and the increased computational cost we decided to keep the results from LEV-4 and not train a deeper network. The LEV-3 network has a similar performance than LEV-4 but at a much lower memory cost so it could be used for cases where there are memory limitations. The similar performance between the two networks may be result of the relatively simple geometry of filaments and walls. In principle the 6 convolutional layers in the encoding path of the LEV-3 model are sufficient to describe local geometry. 

\subsection{Visual inspection}\label{sec:visual}

\begin{figure*}
  \centering
  \includegraphics[width=0.99\textwidth,angle=0.0]{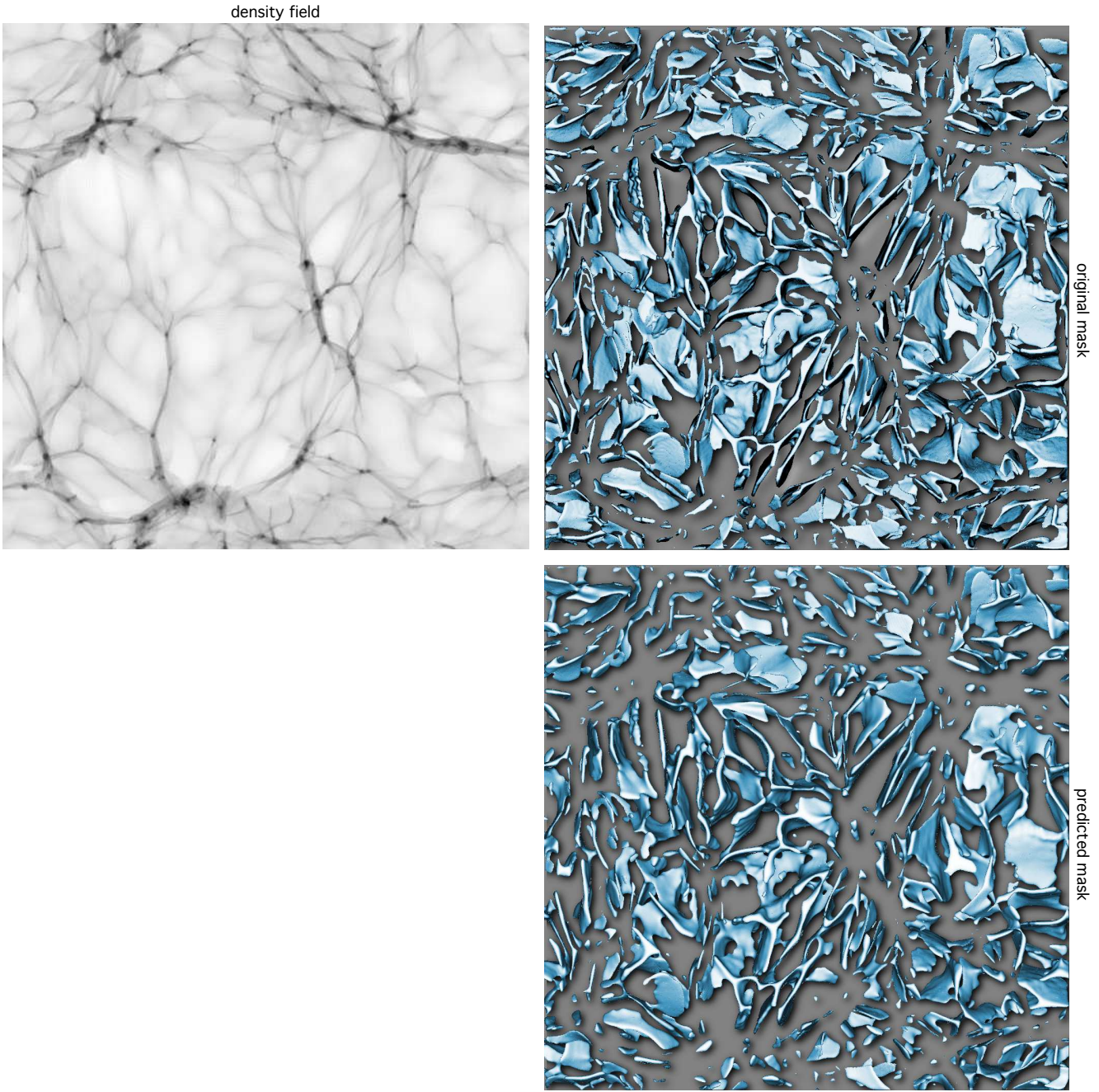}
  \caption{Wall segmentation of the N-body model. Top left: input density field. Top-right: original segmentation masks computed with the MMF method after manually-guided thresholding and cleaning. Bottom-right: predicted segmentation mask computed with the LEV-4 U-Net model. Gray background and shading added for depth cues. The (binary) masks were slightly smoothed for illustration purposes. }\label{fig:64Mpc_prediction_wall}
\end{figure*}

Figures \ref{fig:64Mpc_prediction_fila} and \ref{fig:64Mpc_prediction_wall} shows the predicted filament and wall segmentation masks respectively compared to their corresponding  original segmentation masks and input density field. On first impression both the original and predicted filament and wall masks are visually similar. However, there are more structures in the predicted filament mask compared to the original filament mask. When comparing with the density field we found that several small structures in the predicted mask correspond to tenuous filaments that were not fully or partially identified by the MMF method and thus not included in the original segmentation (see Sec. \ref{sec:enclosed_density}). This shows the generalization ability of the network and it suggests that it can create a more effective internal representation of filaments compared to the feature-engineered manually-guided method used for training.

It is remarkable that both filament and wall predicted  masks do not seem to overlap. Traditional methods based on local geometry tend to confuse both morphologies at their intersections resulting is strong contamination between morphologies in which, for instance,  the wall mask overlaps with the filament mask (see Sec. \ref{sec:MMF}). We measure the degree to which the U-Net predicts a unique solution for filaments and walls by computing their overlapping fraction via the Dice coefficient which has a value of  0.012 confirming our previous observations. For comparison, the raw filament and wall masks from the MMF method (without the cleaning step given by equations \ref{eq:fila_clean} and \ref{eq:wall_clean}) have a Dice coefficient of 0.283. This value is more than one order of magnitude higher than the segmentation computed with the U-Net without any post-processing.

\subsection{Confusion matrix and performance metrics}\label{sec:confusion}

In order to get a quantitative assessment of the ability of the U-Net to reproduce the original segmentation we used the confusion matrix composed of the following terms corresponding to subsets of the segmentation masks:

\begin{itemize}
\item True Positives (TP): voxels inside the original mask and inside the predicted mask.
\item True Negatives (TN): voxels outside the original mask and outside the predicted mask.
\item False Positives (FP): voxels outside the original mask and inside the predicted mask.
\item False Negatives (FN): voxels inside the original mask and outside the predicted mask.\\
\end{itemize}


\begin{table}
\noindent
\renewcommand\arraystretch{1.5}
\setlength\tabcolsep{0pt}
\begin{tabular}{c >{\bfseries}r @{\hspace{0.7em}}c @{\hspace{0.4em}}c @{\hspace{0.7em}}l}
  \multirow{10}{*}{\rotatebox{90}{\parbox{1.1cm}{\bfseries\centering True}}} & 
    & \multicolumn{2}{c}{\bfseries Predicted} & \\
  & & \bfseries P & \bfseries N & \bfseries total \\
  & P & \MyBox{TP}{3,441,239} & \MyBox{FN}{915,631} & 4,356,870 \\[2.4em]
  & N & \MyBox{FP}{1,444,944} & \MyBox{TN}{128,415,912} & 129,860,856 \\
  & total & 4,886,183 & 129,331,543 &
\end{tabular}
\caption{Confusion matrix computed from the filament segmentation.  Columns correspond to the predicted positive ({\bf P}) and negative ({\bf N}) voxels and rows correspond to the true values. The elements of the matrix correspond to the measurements defined in Sec. \ref{sec:confusion}.}
  \label{tab:confusion_filaments}
\end{table}


\begin{table}
\noindent
\renewcommand\arraystretch{1.5}
\setlength\tabcolsep{0pt}
\begin{tabular}{c >{\bfseries}r @{\hspace{0.7em}}c @{\hspace{0.4em}}c @{\hspace{0.7em}}l}
  \multirow{10}{*}{\rotatebox{90}{\parbox{1.1cm}{\bfseries\centering True}}} & 
    & \multicolumn{2}{c}{\bfseries Predicted} & \\
  & & \bfseries P & \bfseries N & \bfseries total \\
  & P & \MyBox{TP}{11,962,370} & \MyBox{FN}{3,404,562} & 15,366,932 \\[2.4em]
  & N & \MyBox{FP}{3,354,966} & \MyBox{TN}{115,495,832} & 118,850,798 \\
  & total & 15,317,336 & 118,900,394 &
\end{tabular}
\caption{Confusion matrix computed from the wall segmentation. Columns correspond to the predicted positive ({\bf P}) and negative ({\bf N}) voxels and rows correspond to the true values. The elements of the matrix correspond to the measurements defined in Sec. \ref{sec:confusion}.}
  \label{tab:confusion_walls}
\end{table}

\noindent Table \ref{tab:confusion_filaments} shows the confusion matrix computed from the filament mask. The model has a very low number of false negatives compared to the true negatives but the true positives are slightly more than twice the number of false positives. This quantifies our observation that the U-Net was able to find more filaments than the MMF method (sec. \ref{sec:visual}). Table \ref{tab:confusion_walls} shows the confusion matrix computed from the wall mask. The number of false negatives is much smaller than the number of true positives indicating that the wall model is better at reproducing the MMF. Walls are simpler geometries (they are the first phase in the Zel'dovich collapse \citep{Zeldovich70}) which perhaps makes them easier to define having only one collapsed axis. 

From the elements in the confusion matrix we computed the following performance metrics:

\begin{equation}\label{eq:accuracy}
\textrm{accuracy} = \frac{\textrm{TP} + \textrm{TN}}{\textrm{TP} + \textrm{TN} + \textrm{FP} + \textrm{FN}},
\end{equation}

\begin{equation}\label{eq:precision}
\textrm{precision} = \frac{\textrm{TP} }{\textrm{TP} +  \textrm{FP}},
\end{equation}

\begin{equation}\label{eq:recall}
\textrm{recall} = \frac{\textrm{TP} }{\textrm{TP} +  \textrm{FN}}.
\end{equation}

\noindent The \textit{accuracy} measures the fraction of correctly classified voxels from all the voxels, \textit{precision} corresponds to the fraction of  correctly classified voxels from all voxels predicted inside the mask and \textit{recall} indicates the fraction of correctly classified voxels from all the voxels in the original mask.  


\begin{table}
  \begin{center}
    \leavevmode
\begin{tabular}{l c c c} 
\hline
\hline
                   & accuracy       & precision  & recall\\
\hline 
filament mask   & 0.98241235 & 0.70427960 & 0.78984202 \\
wall mask          & 0.94963759 & 0.78096935 & 0.77844881 \\
\hline 
\hline
  \end{tabular} 
\caption{Performance metrics for the wall and filament segmentations as defined in equations \ref{eq:accuracy}, \ref{eq:precision} and \ref{eq:recall}.}
  \label{tab:metrics}
  \end{center}
\end{table}

Table \ref{tab:metrics} shows the performance metrics for the predicted  filament and wall masks. The U-Net models have high accuracy for both filaments and walls segmentation. The precision of the filament mask is relatively low (0.7) but this value is underestimated since, as discussed in Sec. \ref{sec:visual}, the U-Net finds filaments nor present in the original mask. The voxels inside such structures are counted as false positives, thus lowering the precision metric below its actual value. The precision of the wall mask is higher than the filament mask as the visual inspection suggests.
The recall values for the filament and wall masks are 0.78 and 0.77 respectively. The recall gives us a better indication of the ability of the model to identify structures in the training dataset independently of false positives. 


\begin{table}
  \begin{center}
    \leavevmode
\begin{tabular}{l c c c} 
\hline
\hline
                   & mean density      & median density\\
\hline 
{\bf MMF:}    &  &  \\
filament   & 11.65 & 2.02 \\
wall          & 0.45  &  0.19\\
\hline 
{\bf U-Net (TP+FP):}    &  &  \\
filament   & 10.84 & 1.78 \\
wall          & 0.43  &  0.18\\
\hline 
{\bf UNet -MMF (FP):}    &  &  \\
filament   & 2.64 & 0.94 \\
filament (eroded)  & 8.82 & 1.42 \\
\hline 
{\bf MMF-UNet (FN):}    &  &  \\
filament   & 2.07 & 0.83 \\
filament (eroded)  & 3.41 & 0.75 \\
\hline 
\hline
  \end{tabular} 
\caption{Mean and median density of voxels inside filament and wall mask computed from the MMF and U-net models. The UNet-MMF column corresponds to the set substraction between the two voxel masks i.e. voxels identified by the U-Net as part of filaments but missed by the MMF (false positives).}
  \label{tab:density}
  \end{center}

\end{table}

%
\subsection{Enclosed density field}\label{sec:enclosed_density}

\begin{figure}
  \centering
  \includegraphics[width=0.49\textwidth,angle=0.0]{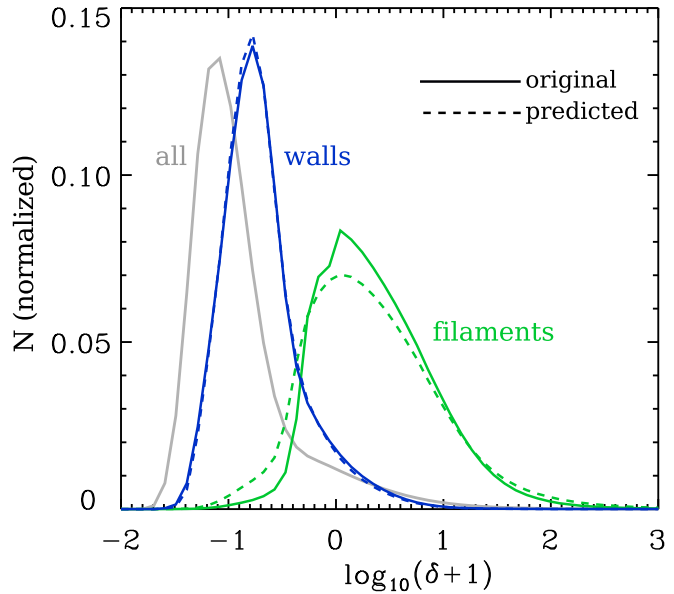}
  \caption{Density distribution of voxels inside filaments (green) and walls (blue) in the original mask (solid line) and the predicted mask (dashed line). The general density distribution is shown in the solid grey line for comparison.}\label{fig:density}
\end{figure}

Figure \ref{fig:density} shows the density distribution inside both original and predicted filament and wall masks. The density distribution of filaments and walls have distinct peaks but also a significant overlap. This is the reason why it is not possible to separate the two morphologies using a density threshold. The density distribution inside the training and predicted  masks are similar for both filaments and walls with the exception of the tails of the density distribution inside filaments where the predicted mask covers slightly more underdense and overdense voxels. 

%
\subsubsection{The nature of the false positive detections}

In order to have a better understanding of the nature of the structures in the FP subset we computed the mean density inside the filament and wall masks (see Table \ref{tab:density}). We excluded voxels in clusters for both masks for consistency as described in section \ref{sec:MMF}. The mean and median density is practically the same for both MMF and U-Net masks. The mean density inside the false positives is also similar for MMF and U-Net masks but when we compare the density at the core of the structures (computed by eroding the filament masks by two voxels) we see that the filaments found by the U-Net and missed from the original mask (FP) have an overdensity of $\delta=8.82$ while the density inside the  false negatives  is $\delta=3.41$. This is a strong indication that these are not noisy underdense filamentary structures but actual overdense coherent structures.

%
\section{Discussion and future work}

We presented the first application of a Deep Convolutional Neural Network  to the semantic segmentation of filaments and walls of an N-body computer simulation.  The performance of the U-Net used here is on par and even produces better segmentations than a current state-of-the-art method while having significantly lower computational cost and being fully automated, requiring one single analysis step after a simple pre-processing. Our model can successfully identify filaments and walls, even finding filamentary structures missed by a carefully hand-tuned method. This has important practical implications as it shows that one can generate high-quality training data with some missing structures and the model will still generate a valid internal representation of the relevant features.

The U-Net presented here was trained to segment filaments and walls following the MMF method. This can be extended to a multi-method approach in which the same model would be trained on segmentations produced by several different methods (in a similar way as large image sets are manually segmented by several people). This way the model will find the common patterns among the different methods, producing a more robust segmentation.

We discussed some guidelines to choose the hyperparameters of the model. This step can be automated by doing a grid search but this is computationally costly and, as shown in Sec. \ref{sec:effect_depth}, the U-Net gives consistent results even with large changes in its architecture (i.e. LEV-3 vs LEV-4). We would expect minor improvements from a hyperparameter grid search and it is beyond the scope of the current work. 

While training times are relatively large (several hours per model) the predictions can be done in a few minutes for a $512^3$ image. The high computational efficiency of deep CNN makes them suitable for analysis of large-volume computer N-body simulations and galaxy surveys as well as large ensemble simulations. Next-generation GPU cards specifically designed for neural network computations with the use of \textit{Tensor Processing Units} (TPU) can speedup computations by a factor of 5 widening even more the execution time gap between deep CNN and traditional methods.

%
\subsection{Future work}

The results presented here can be extended to a fully hierarchical analysis without the need to do any pre-processing such as the initial condition smoothing presented in Sec. \ref{sec:N-body}. One possible way to achieve this would be to introduce several parallel encoding/decoding paths along the same lines as the inception architecture \citep{Szegedy14}. This can perhaps be done at the layer level or as parallel full U-Nets. 

Since there is no restriction in the spatial orientation of filaments and walls the U-Net must not only learn their characteristic geometry but also it must do it for a large range of orientations. This leads to the training of redundant filters that differ only in their orientation. One possible way to address this is to find a rotational invariant representation (which is not clear how to do in 3D) or one could add some layers to rotate the part of the image being processed to a common orientation \citep{Li17}.

In a following paper we will present the more challenging segmentation of a map of galaxies. One advantage of machine learning in this case is that we can create  a mock galaxy catalog and train the model with real-looking data (i.e. redshift distortions, magnitude limits, etc.) but use clean training masks computed before systematic are introduced in the data.

\bibliography{refs} 
\bibliographystyle{mn2e}   

\end{document}